\documentstyle[12pt]{article}

\hoffset =- 20mm
\voffset =- 30mm
\newcommand{\be}[1]{\begin{equation}\label{#1}}
\newcommand{\ee}{\end{equation}}
\newcommand{\ba}[1]{\begin{eqnarray}\label{#1}}
\newcommand{\ea}{\end{eqnarray}}
\newcommand{\nn}{\nonumber}
\newcommand{\nl}{\newline}

\newcommand{\rf}[1]{(\ref{#1})}

\newcommand{\eps}{ \varepsilon }

\newcommand{\R}{ \mbox{\rm I$\!$R} }

\newcommand{\p}{ \mbox{\rm\bf p} }
\newcommand{\z}{ \mbox{\rm\bf z} }

\newcommand{\lorder}{\mbox{ \small $^{<}_{\approx}$ }}

\newcommand{\artanh}{ \mbox{\rm artanh} }
\newcommand{\arcoth}{ \mbox{\rm arcoth} }

\newcommand{\sn}{ \mbox{\rm sn} }

\newcommand{\th}{ \mbox{\rm\tiny th} }
\newcommand{\GH}{ \mbox{\rm\tiny GH} }
\newcommand{\Pl}{ \mbox{\rm\tiny Pl} }
\begin{document}
%
%
\author{ U. Kasper$^1$, M. Rainer$^1$, A. Zhuk$^{1,2}$
\thanks{\noindent Permanent address:
Dept. of Physics, University of Odessa, 2 Petra Velikogo,
270100 Odessa, Ukraine} 
\thanks{\noindent This work was supported by DFG grant 436 RUS 113/7/0} 
}
\title{Integrable Multicomponent Perfect Fluid Multidimensional 
Cosmology II: Scalar Fields.}
\date{}
\maketitle
\vspace*{0.3cm}
\begin{center}
1) Universit\"at Potsdam\\
Institut f\"ur Mathematik\\
Projektgruppe Kosmologie\\
Am Neuen Palais 10\\
PF 601553\\
D-14451 Potsdam\\
Germany
\vspace*{0.3cm}\\ 
2) Freie Universit\"at Berlin\\
Institut f\"ur Theoretische Physik\\
Arnimallee 14\\
D-14195 Berlin\\
Germany\\
\end{center}
\vspace{1cm}
\centerline{\it Received 30 November 1996}
\vspace{1cm}
\abstract
{
\noindent 
We consider anisotropic cosmological models with an universe of dimension 
$4$ or more, factorized into  $n\ge 2$ Ricci-flat spaces, 
containing an $m$-component perfect fluid  
of $m$ non-interacting homogeneous minimally coupled 
scalar fields under special conditions.
We describe the dynamics of the universe:
It has a Kasner-like behaviour near the singularity    
and isotropizes during the expansion to infinity.

Some of the considered models are integrable,
and classical as well as quantum solutions are found.
Some solutions produce inflation from "nothing".
There exist classical asymptotically anti-de Sitter wormholes,
and quantum wormholes with discrete spectrum.
}
%
\nl
\vspace*{0.8cm}
\nl\noindent
{\bf 1. INTRODUCTION}
\setcounter{equation}{0}
\vspace*{0.4cm}
\nl\noindent
It is well known that the isotropic
cosmological model at present time gives a good description of the 
observables part of the universe. On the other hand,
this very fact of our universe's isotropy and homogeneity
is puzzling \cite{Mi}. Even in the papers
which are devoted to the problem of inflation, they start mainly
with the metric of the isotropic Friedmann universe \cite{LLM}.
However, it is possible that at early stages of its evolution the 
universe exhibits an anisotropic behaviour \cite{ZN}.
As it was shown in \cite{LK,BLK}, anisotropic cosmological
models describe the most general approach to the cosmological 
singularity (the initial singularity at some instant $t_0$).
Among anisotropic homogeneous models the Kasner solution \cite{Kas}
represents one of the most simple vacuum solutions of the Einstein 
equations. The Kasner solution is defined on a manifold
\be{1.1}
M = \R \times M_1 \times M_2 \times M_3,
\ee
where the differentiable manifold $M_i$ ($i=1,2,3$) is either $\R$
or $S^1$.

Another very puzzling problem is the fact that the space-time of 
our universe is $4$-dimensional. Fashionable theories of unified
physical interactions (supergravity or superstrings \cite{Wit,DNP,GSW})
use the Kaluza-Klein idea \cite{Kal,Kl} of hidden (or extra) dimensions,
according to which our universe at small (Planckian) scales has 
a dimension more than four. If the extra dimensions are more than
a mathematical construct, we should explain what dynamical processes
lead from a stage with all dimensions developing with the same scale
to the actual stage of the universe, where we have only four external
dimensions and all internal spaces have to be compactified and contracted 
to unobservable scales.

Exploiting these two remarkable ideas (anisotropy and multidimensionality),
it is natural to generalize the manifold \rf{1.1} as follows
\be{1.2}
M = \R \times M_1 \times \dots \times M_n,
\ee
where $M_i$  ($i=1,\ldots,n$) is a $d_i$-dimensional space of constant 
curvature (or, more generally, an Einstein space).
If $n=3$ and $d_1=d_2=d_3=1$ or $n=2$ and $d_1=2$, $d_2=1$ then this
manifold describes an usual anisotropic homogeneous $4$-dimensional
space-time. For $n\geq 2$ and a total dimension $D=1+\sum_{i=1}^{n}d_i>4$
we have an anisotropic multidimensional space-time where one of the spaces
$M_i$ (say $M_1$) describes our $3$-dimensional external space.

Multidimensional cosmological models of the type \rf{1.2}
(with arbitrary $n$) were investigated intensively in the recent decade
(according to our knowledge, starting from the paper \cite{BDLLU}
investigating the stability of the internal spaces).

Quantization  of a multidimensional model with a space-time
\rf{1.2} was first performed in \cite{IMZ}.
Beside vacuum models, there were also cosmological models considered
which contain different types of matter, and exact solutions of
the Einstein equations, and of the Wheeler-De Witt equations 
in the quantum case, were obtained 
(see \cite{33,34} and the extended list of references there).
Exact solutions are of special interest because they can be used for
a detailed study of evolution of the universe
(for example in the approach to the singularity),
of the compactification of the internal spaces, and of the behaviour 
of matter fields.

In the present paper we consider an anisotropic homogeneous
universe of type \rf{1.2}, where all $M_i$ are Ricci-flat.
If $n=3$ and $d_1=d_2=d_3=1$ it describes the usual $4$-dimensional
Bianchi type I model.
We investigate this space-time in the presence of 
$m$ non-interacting minimally coupled scalar fields.
Scalar fields are now popular in cosmology, 
because in most inflationary models the presence of a scalar field
provides homogeneity, isotropy, and almost spatial flatness of the
universe \cite{LLM}.
It was shown in the paper \cite{Zhuk} that for a special form of the scalar
field potentials these scalar fields are equivalent to a $m$-component
perfect fluid.
We exploit this equivalence in \cite{Zhuk} to investigate a two-component
model (a model with $2$ scalar fields). Now we shall integrate this model
in the presence of $3$ scalar fields where one of them is equivalent to 
an ultra-stiff perfect fluid, the second one corresponds to dust,
and the third one is equivalent to vacuum.
The main features of the solutions are the following:
If the parameters of the model permit the universe to run from the
singularity to infinity, then the universe has a Kasner-like behaviour
near the singularity, with isotropization when it goes to infinity.
In the $3$-component integrable case, the universe has de Sitter-like
behaviour in the infinite volume limit.
Superficially, it seems this model is not a good candidate
for a realistic multidimensional cosmology, because of
the isotropization of all directions at late times.
But we shall show that there are particular solutions, which describe a birth
of the universe from "nothing". The parameters of the model in this case
can be chosen in such a way that a scale factor of the external space
undergoes inflation, while the other scale factors remain compactified
near Planck length.
However this model is really only good, if in addition we provide a 
graceful exit mechanism \cite{11}.
For some of the parameters the infinite volume limit takes place
in the Euclidean region which has asymptotically anti-de Sitter
wormhole geometry.
This is in fact possible because in the infinite volume limit
the curvature of the 
spherical spatial sections 
of the anti-de Sitter geometry decays to zero,
whence these spatial sections become asymptotically
Ricci flat like the spacial sections of our solution.

Let us also clarify here that, below we refer only
to the local properties of (anti-)de Sitter space. 
The relationships between different charts of de Sitter space,
with different choices of spacial sections, are examined
e.g. in \cite{Schm,EG}, while \cite{MSchm} recently provided a classification  
of different multidimensional representations of spaces of constant
curvature in arbitrary dimension. For a pure $d+1$-dimensional geometry,
without additional fields, different choices of time, i.e. different
slicings into $d$-dimensional spatial hypersurfaces, should be equivalent
due to general covariance. Note however that, in \cite{KZ} and here,
the geometry hosts in general several additional time-dependent 
but spatially homogenous matter fields, and the 
spatial homogeneity of any such field is generally not preserved
under a change of the slicing.

Another interesting Euclidean solution represents  
an instanton which describes tunnelling between a Kasner-like
universe (a baby universe) and an asymptotically de Sitter universe.
Sewing a number of these instantons may provide the Coleman mechanism
\cite{10} for the vanishing cosmological constant.

Note that our asymptotically de Sitter solutions
are different from the generalized de Sitter solutions
considered in \cite{Zhuk95}. There only one factor space 
(the external one) was Ricci flat, and the curvatures of the other 
(internal) factor spaces were fine tuned with the cosmological
constant.

The previous paper \cite{KZ}  has already considered multidimensional
cosmological models in the presence of a $m$-component perfect fluid.
In the case with one non-Ricci-flat space, say $M_1$,
for $n=2$ and $d_1=2$, $d_2=1$, this model describes a usual $4$-dimensional 
Kantowski-Sachs universe (if $M_1$ has positive constant curvature)
or a Bianchi III universe (if $M_1$ has negative constant curvature).
We also found a $3$-component integrable model, where
the universe has a Kasner-like behaviour near the singularity as in the 
present paper, but there is no isotropization at all. 
All scale factors corresponding to Ricci-flat factor spaces $M_i$
are frozen in the infinite volume limit, 
but the negative curvature space $M_1$ grows in time.
From this point of view, the model does not describe usual
$4$-dimensional space-time, because of the missing isotropization,
but it may be a good candidate for a multidimensional cosmology, 
if all frozen internal scale factors are near Planck scale.
For a positive curvature space $M_1$, the infinite volume limit
takes place in the Euclidean region, which there, in contrast to the 
present paper, has wormhole geometry only w.r.t. the space $M_1$,
and the wormhole is asymptotically flat.

In the present paper we consider homogeneous minimally coupled scalar
fields as a matter source. Usually, only real 
scalar fields are taken. Here we admit also purely imaginary scalar fields. 
Such scalar fields imply a negative sign
at the kinetic term in the Lagrangian.
Such scalar fields may arise after conformal transformation 
of real scalar fields with arbitrary
coupling to gravity \cite{35,Ra,Ra1,36}.
They appear also in the Brans-Dicke theories after
the dimensional reduction from higher dimensional theories
\cite{GSW,AKLO,RZ}.
Also the $C$-field of Hoyle and Narlikar has a negative sign in front of the
kinetic term \cite{40}. 
The authors of \cite{41,42} emphasize 
the need for scalar fields with negative kinetic terms
in multidimensional theories in order to fit the observable data
(see also a discussion of this topic in \cite{43}).
As we will show here,
in the particular case of constant $\varphi$, the imaginary scalar field
is equivalent to a negative cosmological constant
which results in an anti-de Sitter universe.
In what follows we do not exclude the possible existence of imaginary
scalar fields, whence in our paper we consider real as well as imaginary
scalar fields.

This paper is organized as follows. In section 2
we describe our model and get an effective perfect fluid 
Lagrangian, exploiting the equivalence between an $m$-component
perfect fluid and $m$ non-interacting scalar fields
with a special class of potentials.
In section 3, 
we investigate the general dynamics of the universe and
its asymptotic behaviour. 
In section 4, classical solutions for
the integrable 3-component models are obtained. 
Classical wormhole solutions are obtained in section 5
where it is also shown that they are asymptotically anti-de Sitter
wormholes.
Section 6 is devoted to the reconstruction of the scalar field potentials.
Solutions to the quantized models are presented in section 7. 
In section 8 we summarize our results.
\nl
\vspace*{0.8cm}
\nl\noindent
{\bf 2. THE MODEL}
\vspace*{0.4cm}
\nl\noindent
We consider a cosmological model on a multidimensional
manifold \rf{1.2} with a metric
\be{2.1}
g=g_{MN}dx^M\otimes dx^N =-e^{2\gamma(\tau)}d\tau\otimes d\tau + 
\sum_{i=1}^n e^{2\beta^i(\tau)}g^{(i)},
\ee
where, for $i=1,\ldots,n$, 
$g^{(i)}=g^{(i)}_{m_i n_i}dx^{m_i}\otimes dx^{n_i}$,
$m_i,n_i=1,\ldots,d_i$, is the metric form  
of the Ricci-flat factor space $M_i$ dimension $d_i$.
\be{2.2}
R_{m_i n_i}\left[g^{(i)}\right]=0, \quad i = 1,\ldots\,n.
\ee
The action of the model is taken in the form
\be{2.3}
S=\frac{1}{2\kappa^2}\int d^{D}x\sqrt{|g|} R[g]+S_{\varphi}+S_{\GH},
\ee
where $S_{\GH}$ is the standard Gibbons-Hawking boundary term,
$\kappa^2$ is the gravitational coupling constant
in dimension $D=\sum_{i=1}^n d_i +1$, 
and $S_{\varphi}=\sum_{a=1}^m S^{(a)}_{\varphi}$ is the action
of $m$ non-interacting minimally coupled homogeneous scalar fields
\be{2.4}
S^{(a)}_{\varphi}=-\int d^{D}x\sqrt{|g|} 
\left[g^{MN}\partial_M{\varphi^{(a)}}\partial_N{\varphi^{(a)}}
+U^{(a)}(\varphi^{(a)})\right].
\ee
For the metric (\ref{2.1}) the action  (\ref{2.3}) reads
\be{2.5}
S=\mu\int d\tau L_s,
\ee
with the Lagrangian
\be{2.6}
L_s=\frac{1}{2} e^{-\gamma + \gamma_0} 
\left(G_{ij}\dot{\beta}^i \dot{\beta}^j
+\kappa^2 \sum_{a=1}^m (\dot\varphi^{(a)})^2\right)
-  \kappa^2 e^{\gamma + \gamma_0}\sum_{a=1}^m U^{(a)}(\varphi^{(a)}). 
\ee
Here $\gamma_0=\sum_{i=1}^n d_i\beta^i$ and $\mu=\prod_{i=1}^n V_i/\kappa^2$
where $V_i$ is the volume of the finite  Ricci-flat spaces $(M_i,g^{(i)})$. 
The components of the minisuperspace metric read 
\be{2.7}
G_{ij}=d_i \delta_{ij}-d_i d_j.
\ee
As in \cite{Zhuk} we subject the scalar fields to 
the perfect fluid energy-momentum constraints 
\be{2.8}
P^{(a)}= \left( \alpha^{(a)} - 1\right)\rho^{(a)},
\ee
with constants $\alpha^{(a)}$, $a=1,\ldots,m$, and the energy densities
\be{2.9}
\rho^{(a)}\equiv -{T^{(a)}}^0_0 =
\frac{1}{2}e^{-2\gamma}(\dot\varphi^{(a)})^2 + U^{(a)}(\varphi^{(a)})
\ee
and momenta   
\be{2.10}
P^{(a)}\equiv {T^{(a)}}^{M}_M =
\frac{1}{2}e^{-2\gamma}(\dot\varphi^{(a)})^2 - U^{(a)}(\varphi^{(a)}),
\quad M=1,\ldots,D-1\ ,
\ee
according to the Lagrangian \rf{2.6}.
In \cite{Zhuk} it was proved that, for cosmological models with a 
metric (\ref{2.1}), the presence of $m$ non-interacting scalar fields 
satisfying the relations (\ref{2.8}) is equivalent to the presence of
an $m$-component perfect fluid with a Lagrangian
\be{2.11}
L_{\rho}=\frac{1}{2} e^{-\gamma + \gamma_0} 
G_{ij}\dot{\beta}^i \dot{\beta}^j
- \kappa^2 e^{\gamma + \gamma_0} \sum_{a=1}^m \rho^{(a)}, 
\ee
and energy densities of the form
\be{2.12}
\rho^{(a)}=A^{(a)} V^{-\alpha^{(a)}}, \qquad a=1,\ldots,m,
\ee
with constants $A^{(a)}$ and a spatial volume scale
\be{2.13}
V=e^{\gamma_0}=\prod_{i=1}^n a_i^{d_i} 
\ee
defined by the scale factors $a_i=e^{\beta^i}$, $i=1,\ldots,n$.
Note, that the total spatial volume is $V_{tot}=\mu\cdot V$.
The energy density $\rho^{(a)}$ is then connected with the 
pressure $P^{(a)}$ via (\ref{2.8}), and 
equations (\ref{2.9}) and (\ref{2.10}) imply
${\alpha}^{(a)} \rho^{(a)}=e^{-2\gamma}(\dot\varphi^{(a)})^2$.
So, for real scalar fields and positive $\alpha^{(a)}$,
the energy density of the perfect fluid is positive.
But, keeping in mind the possibility of imaginary scalar fields
(see Introduction), for the general model we shall also
consider the case $\rho^{(a)}<0$.
Then, the constants $A^{(a)}$ may have any sign.

Assuming the speed of sound in each component of the 
perfect fluid to be less than the speed of light,
\be{2.14}
-|\rho^{(a)}|\ \leq \ P^{(a)}\ \leq \ |\rho^{(a)}|,\qquad a=1,\ldots,m. 
\ee
With  (\ref{2.8}) this implies the inequalities
\be{2.15}
0\ \leq \ \alpha^{(a)}\ \leq \ 2,\qquad a=1,\ldots,m.
\ee
Note that, with $\rho=\sum_{a=1}^m \rho^{(a)}$ and $P=\sum_{a=1}^m P^{(a)}$, 
the energy dominance condition requires only
$-|\rho|\ \leq \ P\ \leq \ |\rho|$, rather than (\ref{2.14}).
In this paper however, although it might be possible to generalize results
for arbitrary $\alpha^{(a)}$, for simplicity we keep the assumption
(\ref{2.14}) in order to make use of the inequalities (\ref{2.15}). 

Exploiting the mentioned equivalence between scalar fields and perfect fluid,
we investigate the dynamics of the universe via the Euler-Lagrange equations
of (\ref{2.11}), and reconstruct the scalar field potentials 
$U^{(a)}(\varphi^{(a)})$ satisfying the perfect fluid constraint (\ref{2.8}).
\nl
\vspace*{0.8cm}
\nl\noindent
{\bf 3. GENERAL DYNAMICS OF THE UNIVERSE}
\vspace*{0.4cm}
\nl\noindent
In the harmonic time gauge 
$\gamma=\gamma_0=\sum_{i=1}^n d_i\beta^i$ 
(see e.g. \cite{IMZ,Ra}), the Lagrangian (\ref{2.11})
with energy densities (\ref{2.12}) just reads
\be{3.1}
L_{\rho}=\frac{1}{2}  
G_{ij}\dot{\beta}^i \dot{\beta}^j
- \kappa^2 e^{2\gamma_0} \sum_{a=1}^m \rho^{(a)}. 
\ee
Then the corresponding scalar (zero energy) constraint
can be imposed as 
\be{3.2}
\frac{1}{2}  G_{ij}\dot{\beta}^i \dot{\beta}^j
+ \kappa^2 e^{2\gamma_0} \sum_{a=1}^m \rho^{(a)}=0.
\ee
The minisuperspace metric may be diagonalized (see also \cite{IMZ}) to
\be{3.3}
G=\eta_{kl}dz^k\otimes dz^l 
=-dz^0\otimes dz^0 + \sum_{i=1}^{n-1} dz^i\otimes dz^i,
\ee
where 
\ba{3.4}
z^0&=&q^{-1}\sum_{j=1}^{n}d_j\beta^j\ =q^{-1} \ln V,
\\
z^i&=&{\left[\left.d_{i}\right/\Sigma_{i}\Sigma_{i+1}\right]}^{1/2}
\sum_{j=i+1}^{n}
d_j\left(\beta^j-\beta^{i}\right)\ ,
\label{3.5}
\ea
$i=1,\ldots,n-1$, and
\be{3.6}
q:={\left[(D-1) / (D-2)\right]}^{1/2}
,\quad \Sigma_k:=\sum_{i=k}^{n}d_i\ .
\ee
With the aid of these transformations the Lagrangian (\ref{3.1})
and the scalar constraint (\ref{3.2})
can be rewritten as
\be{3.7}
L_{\rho}=\frac{1}{2}  
\eta_{kl}\dot z^k \dot z^l
- \kappa^2 \sum_{a=1}^m A^{(a)} \exp({k^{(a)}q z^0}) , 
\ee
\be{3.8}
\frac{1}{2} \eta_{kl}\dot z^k \dot z^l
+ \kappa^2 \sum_{a=1}^m A^{(a)} \exp({k^{(a)}q z^0}) =0
\ee
respectively. 
Here, $k^{(a)}:=2-\alpha^{(a)}$ ($a=1,\ldots,m$), 
whence the inequalities (\ref{2.15})
for $\alpha^{(a)}$ hold also for $k^{(a)}$,
\be{3.9}
0\ \leq \ k^{(a)}\ \leq \ 2.
\ee
The equations of motion for $z^i$, $i=1,\ldots,n-1$, 
simply read
\be{3.10}
\ddot z^i=0,
\ee
and readily yield
\be{3.11}
z^i=p^i\tau+q^i,
\ee
where $\tau$ is the harmonic time, $p^i$ and $q^i$ are  constants.
Clearly the geometry is real if $p^i$ and $q^i$ are real.
The dynamics of $z^0$ is then given by the scalar constraint
(\ref{3.8}), which may now be written as 
\be{3.12}
-\frac{1}{2} (\dot z^0)^2 + \eps 
+ \kappa^2 \sum_{a=1}^m A^{(a)}  \exp({k^{(a)}q z^0}) =0,
\ee
for a real geometry with
\be{3.13}
\eps:=\frac{1}{2} \sum_{i=1}^{n-1} (p^i)^2 \geq 0.  
\ee
The coordinate transformations  (\ref{3.4}) and (\ref{3.5})
can be written as 
\be{3.14}
z^k=\sum_{i=1}^{n} t^k{}_i \beta^i\ ,\quad k=0,\ldots,n-1, 
\ee
whence the inverse is  given by  
\be{3.15}
\beta^i=\sum_{k=0}^{n-1} \bar t^i{}_k z^k\ ,\quad i=1,\ldots,n.
\ee
For $i=1,\ldots,n$, with $t^0{}_i=d_i/q$ 
and $\bar t^i{}_0=[q(D-2)]^{-1}$ we obtain the scale factors
\be{3.16}
a_i=A_i V^{1/(D-1)} e^{\alpha^i\tau},
\ee
where
\be{3.19}
A_i := e^{\gamma^i},\quad
\gamma^i:=\sum_{l=1}^{n-1} \bar t^i{}_l q^l,\quad
\alpha^i:=\sum_{l=1}^{n-1} \bar t^i{}_l p^l.
\ee
The parameters $\alpha^i$ satisfy the relations
\ba{3.20}
\sum_{i=1}^n d_i\alpha^i=0 ,
\\
\label{3.21}
\sum_{i=1}^n d_i(\alpha^i)^2=\sum_{l=1}^{n-1} (p^l)^2=2\eps\ , 
\ea
and, analogously the parameters $\gamma^i$ fulfil 
\ba{3.22}
\sum_{i=1}^n d_i\gamma^i=0 ,
\\
\label{3.23}
\sum_{i=1}^n d_i(\gamma^i)^2=\sum_{l=1}^{n-1} (q^l)^2 .
\ea
From the definition (\ref{3.19}) and the relation (\ref{3.22})
it follows that
\be{3.24}
\prod_{i=1}^n A_i^{d^i}=1.
\ee
Note that using the constraints (\ref{3.20}) and (\ref{3.24}) 
the equation (\ref{3.16}) yields again (\ref{2.13}), i.e.   
$\prod_{i=1}^n a_i^{d^i}=V$.
Recall that $\tau$ in  (\ref{3.16}) is the harmonic time.
The synchronous and harmonic times are related by 
\be{3.25x}
t=\pm\int e^{\gamma_0} d\tau +t_0=\pm\int V d\tau +t_0.
\ee
The expression (\ref{3.16}) shows that the general model 
does not belong to a class with static internal spaces
(see e.g. \cite{BlZ}), but just for $\eps=0$
( i.e. $\alpha^i=0$, $i=1,\ldots,n$), there is a solution 
\be{3.25}
a_i=A_i V^{1/(D-1)} ,\quad i=1,\ldots,n ,
\ee
which is isotropic.

In order to find the dynamical behaviour of the universe we should
now solve the constraint (\ref{3.12}), i.e. the 
mechanical energy conservation equation 
\be{3.26}
\eps=T+U 
\ee
with kinetic energy
$T:=\frac{1}{2} (\dot z^0)^2$ and
potential $U:=- \kappa^2 \sum_{a=1}^m A^{(a)}  \exp({k^{(a)}q z^0})$.
Depending on the parameters $A^{(a)}$ and their signs,
the potential $U$ may exhibit a rich structure with several extrema,
and a classical Lorentzian trajectory is bound by possible turning points
at $\eps=U$. 
Since the general dynamics is very complex,
we investigate the asymptotic behaviour of our model universe
in the limit of large spatial geometries $V\to\infty$ and
near the singularity $V\to 0$.
Without restriction we suppose now
\be{3.27}
0\leq k^{(1)}<\ldots< k^{(m)}\leq 2 . 
\ee
\nl
\underline{1. Limit $V\to\infty$:} 
In the limit $V\to\infty$ (i.e. $z_0\to\infty$) the term 
$- \kappa^2 A^{(m)}  \exp({k^{(m)}q z^0})$ dominates the potential $U$,
whence, for $k^{(m)}\neq 0$, there are two cases to be distinguished:
\vspace*{0.4cm}

\underline{i) $A^{(m)}>0$:}
Here, for $V\to \infty$, the term $\eps$ may  
be neglected. So, the constraint equation (\ref{3.26})
has the asymptotic solution 
\be{3.28}
e^{q z_0}=V\approx  (2\kappa^2 A^{(m)})^{-1/k^{(m)}} 
({\bar q}|\tau|)^{-2/k^{(m)}} ,
\ee
with $2\bar q:=k^{(m)}q$, where (without restriction) we have chosen initial conditions such that
$V\to \infty$ at $\tau\to 0$, when according to  (\ref{3.16})
the system is subject to an isotropization,
\be{3.29}
a_i \sim V^{1/(D-1)} ,\quad V\to \infty, \quad i=1,\ldots,n\ .
\ee
In this limit the harmonic and synchronous times are connected by
\ba{3.30}
|\tau| &\sim & |t|^{ k^{(m)} / (k^{(m)}-2) }\ , 
\quad  k^{(m)}\neq 2\ ,
\\\label{3.31}
|\tau| &\sim & \exp{(- {\sqrt{ 2\kappa^2 A^{(m)} }} q|t|)}\ , 
\quad  k^{(m)}= 2\ .    
\ea
So the synchronous time evolution of the spatial volume  
is (asymptotically for $t\to\infty$) given by
\ba{3.32}
V &\sim & |t|^{2/\alpha^{(m)}}, \quad  k^{(m)}\neq 2\ ,
\\\label{3.33}
V &\sim & \exp{({\sqrt{2\kappa^2 A^{(m)}}}q|t|)} , \quad  k^{(m)}= 2\ ,    
\ea
with scale factors (according to isotropization)    
\ba{3.34}
a_i \sim |t|^{2/\alpha^{(m)}(D-1)}\ , 
\quad  k^{(m)}\neq 2\ ,
\\
\label{3.35}
a_i \sim \exp{(\frac{\sqrt{2\kappa^2 A^{(m)}}q}{D-1}|t|)}\ , 
\quad  k^{(m)}= 2\ .    
\ea
Taking a usual anisotropic space-time model
($D=4$, $n=3$, $d_1=d_2=d_3=1$) then for large (synchronous) times
the formulas (\ref{3.34}) and (\ref{3.35}) yield scale factors
$a_i \sim |t|^{2/3}$ 
for $k^{(m)}=1$ (dust) 
and $a_i \sim \exp{({\sqrt{\kappa^2 A^{(m)}/3}}|t|)}$
for  $k^{(m)}=2$ (vacuum).
Asymptotically, power-law inflation (with power $p>1$) takes place
for $0<\alpha^{(m)}<2/(D-1)$, and $\alpha^{(m)}=2/(D-1)$
yields a generalized Milne universe.
\vspace*{0.4cm}

\underline{ii) $A^{(m)}<0$:}
Here, the Lorentzian region has a boundary at the turning point  $V_{\max}$ 
of the volume scale, which in the large energy limit $\eps\to\infty$
is asymptotically given as 
\be{3.36}
V_{\max}\approx  \left[\frac{\eps}{\kappa^2 |A^{(m)}|}\right]^{1/k^{(m)}}\ .
\ee
The region with $V>V_{\max}$ is the Euclidean sector.
For $V>>V_{\max}$, we obtain the
asymptotically isotropic solution
\be{3.37}
a_i\sim V^{1/(D-1)}  \approx  
\left[{\sqrt{2\kappa^2 |A^{(m)}|}}\bar q|\tau|\right]^{-2/k^{(m)}(D-1)} \ .
\ee
In the Euclidean region, we obtain a classical wormhole
w.r.t. each factor space. 
With constants of integration (in \rf{3.11}) $p_i=0$ (i.e. $\alpha^i=0$), 
$i=1,\ldots,n$, the wormhole takes its most simple and symmetric form. 
Then the throats are given by
\be{3.38}
a_{(\th)i}\approx A_i  
\left[\eps/\kappa^2 |A^{(m)}|\right]^{1/k^{(m)}(D-1)} \ .
\ee
In the case $k^{(m)}=2$ we obtain asymptotically (for $t\to\infty$) 
anti-de Sitter wormholes with synchronous time scale factors 
\be{3.39}
a_i \sim \exp{(\frac{\sqrt{2\kappa^2 |A^{(m)}|}q}{D-1}|t|)}\ ,\quad
i=1,\ldots,n\ . 
\ee
\nl
\underline{2. Limit $V\to 0$:} 
For $k^{(1)}\neq 0$, 
in the small volume limit $V\to 0$, i.e. $z^0\to-\infty$,
the potential vanishes $U\to 0$.
So, for $\eps>0$, 
we obtain (asymptotically for $t\to 0$) a 
(multidimensional) Kasner universe \cite{BlZ1,Iv}, with scale factors  
\be{3.40}
a_i \sim |t|^{\bar\alpha^i}\ ,\quad
i=1,\ldots,n\ . 
\ee
with parameters $\bar{\alpha^i}$ satisfying 
\be{3.41}
\sum_{i=1}^n d_i{\bar{\alpha^i}}=1\ ,\quad
\sum_{i=1}^n d_i{(\bar{\alpha^i})^2}=1\ .
\ee
If $k^{(1)}=0$, then $U\to -\kappa^2 A^{(1)}$ for $z^0\to-\infty$.
Here, for $E:=\eps+\kappa^2 A^{(1)}>0$,  
we obtain (asymptotically for $t\to 0$) a 
generalized Kasner universe \cite{BlZ1}, i.e. scale factors  
(\ref{3.40})
with parameters $\bar{\alpha^i}$ satisfying 
\be{3.42}
\sum_{i=1}^n d_i \bar{\alpha^i}=1\ ,\quad
\sum_{i=1}^n d_i{(\bar{\alpha^i})^2}=1-\bar\alpha^2 ,
\ee
with the parameter $\bar\alpha\to 0$ for $A^{(1)}\to 0$. 

In the exceptional case $E=\eps+\kappa^2 A^{(1)}=0$
the term of the matter component $a=2$ dominates the constraint 
(\ref{3.12}), whence we obtain (compare also \cite{Zhuk})
\ba{3.43}
V&\sim& t^{2/{\alpha^{(2)}} }\ ,
\\
\label{3.44}
a_i&\sim& t^{2/[(D-1){\alpha^{(2)}}]}\exp {\left\{\alpha^i
f(\alpha^{(2)}){t^{-\frac{2-{\alpha^{(2)}}}{\alpha^{(2)}} }}\right\}} , \quad 
i=1,\ldots ,n\ ,
\ea
where $f(x):={\left(\frac{2-{x} }{{x} }\right)}^{
\frac{(2-{x} )}{x}}
{\left[\frac{2}{{(2-{x} )}^2 q^2\kappa^2 A^{(2)} }\right]}^{
\frac{1}{x}}$.
In another exceptional case where $\eps=0$ 
(i.e. $\alpha^i=0$, $i=1,\ldots,n$) the universe is isotropic
everywhere, i.e. $a_i\sim V^{1/{D-1}}$, $i=1,\ldots,n$.
If, for example, $k^{(1)}=0$ (and $A^{(1)}=0$) 
we obtain from (\ref{3.43}) or (\ref{3.44}) 
\be{3.45}
a_i\sim t^{2/{[(D-1)\alpha^{(2)}]} }\ .
\ee
\nl
\vspace*{0.8cm}
\nl\noindent
{\bf 4. INTEGRABLE $3$-COMPONENT MODEL. CLASSICAL SOLUTIONS}
\vspace*{0.4cm}
\nl\noindent
In this chapter, we consider the integrable case of a three-component 
perfect fluid ($m=3$) where one of them ($a=1$) is ultra-stiff matter  
$(k^{(1)}=0, 
\alpha^{(1)}=2)$, the second one ($a=2$) is dust $(k^{(2)}=1, 
\alpha^{(2)}=1)$, and the third one ($a=3$) is vacuum $(k^{(3)}=2, 
\alpha^{(3)}=0)$. The case $k^{(1)}=0, k^{(3)}=2k^{(2)}$ with $0<k^{(2)}\le 
2$ is also 
included if one substitutes $q$ by $\bar{q}=k^{(2)}q$.

The constraint equation \rf{3.12} reads in this case
\begin{equation}\label{4.1}
-\frac{1}{2}\left(\dot{z}^0\right)^2+\eps 
+\kappa^2A^{(1)}+\kappa^2A^{(2)}e^{qz^0}+\kappa^2A^{(3)}e^{2qz^0}=0
\end{equation}
and can be rewritten like
\begin{equation}\label{4.2}
E=\frac{1}{2}\left(\dot{z}^0\right)^2+U(z^0)\, ,
\end{equation}
where
\begin{equation}\label{4.3}
E:=\eps+\kappa^2A^{(1)}
\end{equation}
and the potential $U(z^0)$ is
\begin{equation}\label{4.4}
U(z^0):=-Be^{qz^0}-Ce^{2qz^0}
\end{equation}
with the definitions $B:=\kappa^2A^{(2)}$ and $C:=\kappa^2A^{(3)}$.

As mentioned in the introduction, for a complete description of the model 
the parameters E, B, and C are considered to take positive and negative 
values. Then, we have four qualitatively different shapes of the potential 
\rf{4.4} (see Fig. 1 and Fig. 2). For each of them, we shall solve the 
constraint equation separately. Eq. \rf{4.2} integrates to
\begin{equation}\label{4.5}
\int{\frac{dV}{V\sqrt{E+BV+CV^2}}}=\pm \sqrt{2}q(\tau -\tau_0)\, ,
\end{equation}
where $\tau$ is the harmonic time coordinate, and $\tau_0$ is a constant of 
integration.
\nl\nl
\noindent
\underline{i) $B>0$, $C>0$ (see Fig. 1):}
The solutions of equations \rf{4.5} are
\begin{eqnarray}\label{4.6}
V=\frac{1}{B}\frac{1}{\frac{q^2}{2}\left[f^2-\frac{2C}{q^2B^2}\right]},
&& E=0,
\\\label{4.7}
V=\frac{4Ef}{(B-f)^2-4EC}, 
&& E>0,
\\\label{4.8}
V=\frac{2|E|}{B+\sqrt{|\Delta|}f}, 
&& E<0,
\end{eqnarray}
where $\Delta:=4EC-B^2$ ($|\Delta|=B^2+4|E|C$ for $E<0$) and
\begin{eqnarray}
\label{4.9}
f=\tau -\tau_0, &E=0,& \frac{\sqrt{2C}}{qB}\le 
|\tau-\tau_0|<+\infty \, ,
\\\label{4.10}
f=\exp{\left(\sqrt{2E}q(\tau-\tau_0)\right)}, &E>0,& 
\ln{\left(B+2\sqrt{EC}\right)}\le\ln{f}<+\infty \, ,
\\\label{4.11}
f=\sin\left(\sqrt{2|E|}q(\tau-\tau_0)\right), &E<0,& -\arcsin \left(
\frac{B}{\sqrt{|\Delta|}} \right)\le  \arcsin f\le \frac{\pi}{2}\, .
\end{eqnarray}
The synchronous and harmonic time coordinate are related via
\begin{eqnarray}\label{4.12}
\tau-\tau_0=\frac{\sqrt{2C}}{qB}\coth\left(\sqrt{\frac{C}{2}}qt\right),&E=0,&
\\\label{4.13}
\exp\left(\sqrt{2E}q(\tau-\tau_0)\right)=B+\sqrt{4EC}
\coth\left(\sqrt{\frac{C}{2}}qt\right),&E>0,&
\\\label{4.14}
\tan\left(\sqrt{|E|/2}q(\tau-\tau_0)\right)=\frac{\sqrt{|\Delta|}}{B}
\left[\sqrt{\frac{4|E|C}{|\Delta|}}
\coth\left(\sqrt{\frac{C}{2}}qt\right)-1\right],&E<0.&
\end{eqnarray}
Using these relations, we obtain the expressions for the volume of the 
universe in synchronous time:
\begin{eqnarray}\label{4.15}
V=\frac{B}{C}\sinh^2\left(\sqrt{\frac{C}{2}}qt\right), &E=0,& 
|t|<\infty,
\\\label{4.16}
V=\frac{1}{C}
\left[B+\sqrt{4EC}\coth\left(\sqrt{\frac{C}{2}}qt\right)\right]
\sinh^2\left(\sqrt{\frac{C}{2}}qt\right),& E>0,&
0\le t<+\infty,
\\\label{4.17}
V=\frac{2|E|(1+\tan^2(y/2))}{B(1+\tan^2(y/2))+2\sqrt{|\Delta|}\tan(y/2)}, 
&E<0 ,&
\end{eqnarray}
where $\tan(y/2)=\tan\left(\sqrt{|E|/2}q(\tau-\tau_0)\right)$ is given by 
\rf{4.14}. 
Expression \rf{4.17} can be written in a more convenient way if the 
parameter $\tau_0$ is chosen such that equation \rf{4.8} is symmetric with 
respect to the turning point $V_0=\left(-B+\sqrt{|\Delta|}\right)/2C$, namely
\begin{equation}\label{4.18}
V=\frac{2|E|}{B+\sqrt{|\Delta|}\cos\left(\sqrt{2|E|}q\tau\right)}, \quad
|\tau|<\frac{1}{\sqrt{2|E|q}}\left[\frac{\pi}{2}+
\arcsin\left(\frac{B}{\sqrt{|\Delta|}}\right)\right].
\end{equation}
In this case,
\begin{equation}\label{4.19}
\tan\left(\sqrt{|E|/2}q\tau\right)=\frac{\sqrt{4|E|C}}{\sqrt{|\Delta|}-B}
\tanh\left(\sqrt{\frac{C}{2}}qt\right)
\end{equation}
and for the volume results
\begin{equation}\label{4.20}
V=\frac{1}{2C}\left[\sqrt{|\Delta|}-B+
\left(\sqrt{|\Delta|}+B\right)\tanh^2\left(\sqrt{\frac{C}{2}}qt\right)\right]
\cosh^2\left(\sqrt{\frac{C}{2}}qt\right), 
\quad 
|t|<\infty.
\end{equation}
The region $V<V_0$ is the Euclidean sector and we obtain the instanton by 
analytic continuation $t\rightarrow -it$ in formula \rf{4.20}:
\begin{equation}\label{4.21}
V=\frac{1}{2C}\left[\sqrt{|\Delta|}-B
-\left(\sqrt{|\Delta|}+B\right)\tan^2\left(\sqrt{\frac{C}{2}}qt\right)\right]
\cos^2\left(\sqrt{\frac{C}{2}}qt\right)
\end{equation}
with $|t|\le\frac{2}{\sqrt{2C}q}\arctan\sqrt{\frac{\sqrt{|\Delta|}-B}
{\sqrt{|\Delta|}+B}}.$

On the quantum level, this instanton is responsible for the birth of the 
universe from ``nothing".
\nl\nl
\noindent
\underline{ii) $B<0$, $C>0$ (see Fig. 2):}
In this case, the maximal value of the potential $U(z^0)$ is $U_m=B^2/4C$ at 
$z_m^0=\frac{1}{q}\ln\left(|B|/2C\right)$ and for $0<E<V_m$ we have two 
turning points, namely:
\begin{equation}\label{4.22}
V_0^{(1,2)}=\left(|B|\pm\sqrt{|\Delta|}\right)/2C\, ,
\end{equation}
where $|\Delta|=B^2-4EC$. Classical motion takes place either for 
$0\le V\le V_0^{(1)}$ or for $V_0^{(2)}\le V<+\infty$.

If $E\le 0$, we have one turning point only, namely
\begin{equation}\label{4.23}
V_0=|B|/C, \quad E=0,
\end{equation}
\begin{equation}\label{4.24}
V_0=\left(|B|+\sqrt{|\Delta|}\right)/2C,\quad E<0\, ,
\end{equation}
where $|\Delta|=B^2+4|E|C$ and classical motion takes place for $V\ge V_0$.

The solutions of the equation \rf{4.5} read
\begin{eqnarray}\label{4.25}
V=\frac{1}{|B|}\frac{1}{\frac{q^2}{2}\left(\frac{2C}{B^2q^2}-f^2\right)}, 
&& E=0,
\\\label{4.26}
V=\frac{4Ef}{(|B|+f)^2-4EC}, 
&& 0<E<U_m,\quad 0\le V\le V_0^{(1)}, 
\\\label{4.27}
V=\frac{4Ef}{4EC-(|B|-f)^2}, 
&& 0<E<U_m,\quad V_0^{(2)}\le V<+\infty, 
\\\label{4.28}
V=\frac{4Ef}{(|B|+f)^2-4EC},
&& E>U_m,
\\\label{4.29}
V=\frac{2|E|}{\sqrt{|\Delta|}f-|B|},
&& E<0,
\end{eqnarray}
where
\begin{eqnarray}\label{4.30}
f=\tau-\tau_0, 
&& E=0, 
\\\nn 
&& |\tau-\tau_0|<\sqrt{2C}/q|B|,
\\\label{4.31}
f=\exp\left(\sqrt{2E}q(\tau-\tau_0)\right),
&& 0<E<U_m,\quad V\le V_0^{(1)},
\\\nn 
&& \ln\sqrt{|\Delta|}\le \ln f<+\infty,
\\\label{4.32}
f=\exp\left(\sqrt{2E}q(\tau-\tau_0)\right),
&& 0<E<U_m,\quad V\ge V_0^{(2)},
\\\nonumber
&& \ln\left(|B|-\sqrt{4EC}\right)\le \ln f\le 
\ln\sqrt{|\Delta|},
\\\label{4.33}
f=\exp\left(\sqrt{2E}q(\tau-\tau_0)\right),
&& E>U_m,
\\\nonumber
&& \ln\left(-|B|+\sqrt{4EC}\right)\le \ln f<+\infty,
\\\label{4.34} 
f=\sin\left(\sqrt{2|E|}q(\tau-\tau_0)\right),
&& E<0,
\\\nonumber
&& \arcsin\frac{|B|}{\sqrt{|\Delta|}}\le \arcsin f\le \frac{\pi}{2}.
\end{eqnarray}
The harmonic and synchronous time coordinates are related via
\begin{eqnarray}\label{4.35}
\tau=\frac{2C}{|B|q}\tanh\left(\sqrt{\frac{C}{2}}qt\right), 
&& E=0,
\\\label{4.36}
f=\sqrt{4EC}\cot\left(\sqrt{\frac{C}{2}}qt\right)-|B|,
&& 0<E<B^2/4C,\quad V<V_0^{(1)},
\\\label{4.37}
f=-\sqrt{4EC}\tanh\left(\sqrt{\frac{C}{2}}qt\right)+|B|, 
&& 0<E<B^2/4C,\quad V>V_0^{(2)},
\\\label{4.38}
f=\sqrt{4EC}\coth\left(\sqrt{\frac{C}{2}}qt\right)-|B|,
&& E>B^2/4C,
\\\label{4.39}
\tan\left(\sqrt{|E|/2}q\tau\right)=\frac{\sqrt{4|E|C}}{\sqrt{|\Delta|}+|B|}
\tanh\left(\sqrt{\frac{C}{2}}qt\right),
&& E<0,
\end{eqnarray}
where in \rf{4.35} and \rf{4.39} 
the constant $\tau_0$ is chosen 
such that the expressions are symmetric with respect to the turning point at 
the minimum. Then, the volume of the universe is
\begin{eqnarray}\label{4.40}
V&=&\frac{|B|}{C}\cosh^2\left(\sqrt{\frac{C}{2}}qt\right),\quad 
E=0,\quad |t|<+\infty,
\\
\label{4.41}
V&=&\frac{1}{C}\left[\sqrt{4EC}\coth
\left(\sqrt{\frac{C}{2}}qt\right)-|B|\right]
\sinh^2\left(\sqrt{\frac{C}{2}}qt\right),\quad  
\\
&&0<E<B^2/4C,\quad V<V_0^{(1)},\quad 0\le t\le 
\frac{2}{\sqrt{2C}q}\arcoth\frac{|B|+\sqrt{|\Delta|}}{\sqrt{4EC}},\nonumber\\
\label{4.42}
V&=&\frac{1}{2C}
\left[|B|+\sqrt{|\Delta|}-\left(|B|-\sqrt{|\Delta|}\right)\tanh^2
\left(\sqrt{\frac{C}{2}}qt\right)\right]
\cosh^2\left(\sqrt{\frac{C}{2}}qt\right),\quad 
\\
&&0<E<B^2/4C,\quad 
V>V_0^{(2)},\quad |t|<+\infty 
\nonumber\\
V&=&\frac{1}{C}\left[\sqrt{4EC}\coth
\left(\sqrt{\frac{C}{2}}qt\right)-|B|\right]
\sinh^2\left(\sqrt{\frac{C}{2}}qt\right),
\\%
&& E>B^2/4C,\quad 0\le t<+\infty 
\nonumber
\\\label{4.44}
V&=&\frac{1}{2C}\left[\sqrt{|\Delta|}+|B|
+\left(\sqrt|\Delta|-|B|\right)\tanh^2\left(\sqrt{\frac{C}{2}}qt\right)\right]
\cosh^2\left(\sqrt{\frac{C}{2}}qt\right),
\\
&& 
E<0,\quad 
|t|<+\infty 
\nonumber .
\end{eqnarray}
Eqs. \rf{4.40}, \rf{4.42}, and \rf{4.44} are written in a symmetric 
way with respect to the turning point at $t=0$. The instanton 
solutions can be obtained by analytic continuation of these 
symmetric expressions and result in:
\begin{eqnarray}\label{4.45}
V&=&\frac{|B|}{C}\cos^2\left(\sqrt{\frac{C}{2}}qt\right), \quad E=0,\quad
|t|\le \pi/\sqrt{2C}q,
\\\label{4.46}
V&=&\frac{1}{2C}\left[|B|+\sqrt{|\Delta|}
+\left(|B|-\sqrt{|\Delta|}\right)\tan^2\left(\sqrt{\frac{C}{2}}qt\right)\right]
\cos^2\left(\sqrt{\frac{C}{2}}qt\right),\\
&&0<E<B^2/4C\quad (|\Delta|=B^2-4EC),\quad |t|\le \pi/\sqrt{2C}q,\nonumber\\
V&=&\frac{1}{2C}\left[|B|+\sqrt{|\Delta|}
+\left(|B|-\sqrt{|\Delta|}\right)\tan^2\left(\sqrt{\frac{C}{2}}qt\right)\right]
\cos^2\left(\sqrt{\frac{C}{2}}qt\right),
\label{4.47}\\
&&E<0 \quad (|\Delta|=B^2+4|E|C),\quad 
|t|\le \frac{2}{\sqrt{2C}q}\arctan
\sqrt{\frac{\sqrt{|\Delta|}+|B|}{\sqrt{|\Delta|}-|B|}}
\nonumber.
\end{eqnarray}
In equation \rf{4.46}, the instanton is symmetric with respect 
to the turning point $V_0^{(2)}$. For the same instanton but 
now symmetric with respect to the turning point at 
$V_0^{(1)}$, we have
\begin{eqnarray}\label{4.48}
V&=&\frac{1}{2C}\left[|B|-
\sqrt{|\Delta|}+\left(|B|+\sqrt{|\Delta|}\right)\tan^2
\left(\sqrt{\frac{C}{2}}qt\right)\right]
\cos^2\left(\sqrt{\frac{C}{2}}qt\right),
\\\nn
&&0<E<B^2/4C,\quad |t|\le \pi/\sqrt{2C}q.
\end{eqnarray}
All the instantons \rf{4.45} to \rf{4.48} 
are responsible on the quantum level for 
the birth of the universe from ``nothing". The instanton \rf{4.46}, 
\rf{4.48} is of 
special interest. Its qualitative shape is seen in Fig. 3 where 
$V_{min}=V_0^{(1)}$ and $V_{max}=V_0^{(2)}$. The instanton describes 
tunnelling between a multidimensional Kasner-like universe (a baby universe) 
and a multidimensional de Sitter universe because, as was mentioned in 
chapter 3 and as we shall demonstrate more precisely latter, the limit 
$V\rightarrow 0$ corresponds to a Kasner-like universe and in the limit 
$V\rightarrow \infty$ we obtain an (isotropic) de Sitter universe 
(in \cite{6} to \cite{9} analogous types of an instanton describing tunnelling between a Friedmann 
universe
and a de Sitter universe were obtained for a different 
model). The instanton may also represent the birth (demise) 
of a de Sitter universe (see \rf{4.46}) and a baby universe 
(see \rf{4.48}) from (into) ``nothing". As was demonstrated 
in \cite{6,7}, the instanton may be extended beyond $V=V_{min, 
max}$ gluing together a number of Euclidean manifolds (see 
Fig. 4). Such gluing may provide the Coleman mechanism \cite{10} 
that establishes  the vanishing of the cosmological 
constant.

The case with $E=B^2/4C$ is degenerated. Here, two turning 
points coincide with each other: $V_0^{(1)}=V_0^{(2)}\equiv 
V_0=|B|/2C$. In this case, we obtain with the synchronous 
time coordinate $t$ for the two volumes the expressions
\begin{equation}\label{4.49}
V=\frac{2E}{|B|}\left[1-\exp\left(-\frac{|B|}{\sqrt{2E}}qt\right)\right],\quad
0\le t<+\infty,
\end{equation}
which describes an infinitely long lasting rolling down from 
the unstable equilibrium position $V_0$ to the singularity 
$V=0$ and
\begin{equation}\label{4.50}
V=\frac{2E}{|B|}\left[1+\exp\left(\frac{|B|}{\sqrt{2E}}qt\right)\right],\quad
|t|<+\infty
\end{equation}
describing the infinitely long lasting rolling down with 
$V\rightarrow \infty$.

To obtain solutions in the two remaining cases $B>0,\quad 
C<0$ (see Fig. 2) and $B<0,\quad C<0$ (see Fig. 1), it is not 
necessary to solve equation \rf{4.5} again. We can instead take the 
solutions found already in subsections i) and ii). It is 
clear that the Euclidean solutions obtained there are 
Lorentzian ones here and vice versa Lorentzian solutions of 
subsections i) and ii) are Euclidean ones here. What we have 
to do is the evident substitutions $B\rightarrow |B|$,  
$C\rightarrow |C|$, and $E\rightarrow -E$ where it is necessary. For example:
\nl\nl
\noindent
\underline{iii) $B>0$, $C<0$:}
From \rf{4.45}, \rf{4.47}, and \rf{4.48}, we obtain respectively
\begin{eqnarray}\label{4.51}
V&=&\frac{B}{|C|}\cos^2\left(\sqrt{|C|/2}qt\right),\quad 
E=0,\quad |t|\le \pi/q\sqrt{2|C|},
\\\label{4.52}
V&=&\frac{1}{2|C|}
\left[\sqrt{|\Delta|}+B-
\left(\sqrt{|\Delta|}-B\right)\tan^2\left(\sqrt{|C|/2}qt\right)\right]
\cos^2\left(\sqrt{|C|/2}qt\right),
\\
&&E>0,\quad |t|\le 
\frac{2}{\sqrt{2|C|}q}\arctan\left[\left(\sqrt{|\Delta|}+B\right)/
\left(\sqrt{|\Delta|}-B\right)\right]^{1/2},
\nonumber\\
\label{4.53}
V&=&\frac{1}{2|C|}
\left[B-\sqrt{|\Delta|}+
\left(B+\sqrt{|\Delta|}\right)\tan^2\left(\sqrt{|C|/2}qt\right)\right]
\cos^2\left(\sqrt{|C|/2}qt\right),
\\%
&&-B^2/4|C|<E<0,\quad |t|<\pi/\sqrt{2|C|}q\, \nonumber .
\end{eqnarray}
Solution \rf{4.52} is symmetric with respect to the classical 
turning point. To investigate the limit $|t|\rightarrow 0$ 
it is better to give another representation of the same 
solution, namely
\begin{eqnarray}\label{4.54}
V&=&\frac{1}{|C|}\left[B+\sqrt{4E|C|}
\cot\left(\sqrt{|C|/2}qt\right)\right]
\sin^2\left(\sqrt{|C|/2}qt\right),
\\\nonumber
&&E>0,\quad 0\le t\le 
\frac{2}{\sqrt{2|C|}q}\arctan\frac{\sqrt{|\Delta|}-B}{\sqrt{4E|C|}}.
\end{eqnarray}
In this case, the harmonic time coordinate and the synchronous one are 
related via
\begin{equation}\label{4.55}
\exp\left(\sqrt{2E}q(\tau-\tau_0)\right)=B+\sqrt{4E|C|}\cot
\left(\sqrt{|C|/2}qt\right) .
\end{equation}
Equation \rf{4.53} is symmetric with respect to the turning 
point $V_0^{(1)}=\left(B-\sqrt{|\Delta|}\right)/2|C|$. Its 
analytic continuation gives a ``parent instanton" 
(see \rf{4.41}) with
\begin{eqnarray}\label{4.56}
V&=&\frac{1}{|C|}\left[\sqrt{4EC}\coth
\left(\sqrt{|C|/2}qt\right)-B\right]
\sinh^2\left(\sqrt{|C|/2}qt\right),
\\\nn
&&-B^2/4|C|<E<0,\quad 0\le t\le 
\frac{2}{\sqrt{2|C|}q}\arcoth\frac{B+\sqrt{|\Delta|}}{\sqrt{4EC}},
\end{eqnarray}
which is responsible for the birth of a baby universe from 
``nothing". The Lorentzian solution \rf{4.53} symmetrically 
written with respect to the turning point 
$V_0^{(2)}=\left(B+\sqrt{|\Delta|}\right)/2|C|$ reads  
(see \rf{4.46})
\begin{eqnarray}\label{4.57}
V&=&\frac{1}{2|C|}\left[B+\sqrt{|\Delta|}
+\left(B-\sqrt{|\Delta|}\right)\tan^2\left(\sqrt{|C|/2}qt\right)\right]
\cos^2\left(\sqrt{|C|/2}qt\right),
\\\nn
&&-B^2/4|C|<E<0,\quad |t|\le \pi/\sqrt{2|C|}q.
\end{eqnarray}
\nl\nl
\noindent
\underline{iv) $B<0$, $C<0$:}
Here, a Lorentzian region exists for $E>0$ only. From 
equation \rf{4.5}, we obtain
\begin{eqnarray}\label{4.58}
V&=&\frac{1}{|C|}
\left[4E|C|\cot\left(\sqrt{|C|/2}qt\right)-|B|\right]
\sin^2\left(\sqrt{|C|/2}qt\right),
\\\nn
&&E>0,\quad 0\le t\le 
\frac{2}{\sqrt{2|C|}q}\mbox{arccot}\frac{\sqrt{|\Delta|}+|B|}{4E|C|}
\end{eqnarray}
and the equation relating the harmonic time coordinate and 
the synchronous one reads
\begin{equation}\label{4.59}
\exp\left(\sqrt{2E}q(\tau-\tau_0)\right)=\sqrt{4E|C|}
\cot\left(\sqrt{|C|/2}qt\right)-|B|.
\end{equation}
These equations are useful for the investigation of the small 
time limit $|t|\rightarrow 0$. 

To obtain an instanton solution (wormhole) it is necessary to rewrite 
equation \rf{4.58} symmetrically with respect to the classical 
turning point $V_0=\left(-|B|+\sqrt{|\Delta|}\right)/2|C|$. 
We can reformulate equation \rf{4.58} or use directly equation 
\rf{4.21}. The result is
\begin{eqnarray}\label{4.60}
V&=&\frac{1}{2|C|}
\left[\sqrt{|\Delta|}-|B|-\left(\sqrt{|\Delta|}+|B|\right)
\tan^2\left(\sqrt{|C|/2}qt\right)\right]
\cos^2\left(\sqrt{|C|/2}qt\right),
\\
&&E>0,\quad |t|\le 
\frac{2}{\sqrt{2|C|}q}\arctan\sqrt\frac{\sqrt{|\Delta|}-|B|}
{\sqrt{|\Delta|}+|B|}.\nonumber
\end{eqnarray}

We shall investigate now the small time limit 
$|t|\rightarrow 0$ for Lorentzian solutions. As we shall see, 
it corresponds to the vanishing volume limit $V\rightarrow 
0$ and takes place for $E\ge 0$ if $B, C>0$ or $B>0, C<0$ 
and for $E>0$ if $B, C<0$ or $B<0, C>0$ (see Fig. 1 and 
Fig. 2). First, we consider the case of positive energies 
$E>0$. As follows from \rf{4.13}, \rf{4.36}, \rf{4.55}, and \rf{4.59}
\begin{equation}\label{4.61}
\exp\left(\sqrt{2E}q\tau\right)\sim t^{-1},\quad t\rightarrow 0
\end{equation}
and with the help of equations \rf{3.16}, \rf{3.20}, and \rf{3.21} we 
obtain for the scale factors in this limit the expressions
\begin{equation}\label{4.62}
a_i\approx \bar{A}_it^{\bar{\alpha}^i}, \quad t\rightarrow
0,
\end{equation}
where 
\begin{equation}\label{4.63}
\bar{\alpha}^i=\frac{1}{D-1}-\frac{1}{\sqrt{2E}q}\alpha^i
\end{equation}
and the parameters satisfy the conditions
\begin{equation}\label{4.64}
\sum_{i=1}^n d_i\bar{\alpha}^i=1,
\end{equation}
\begin{equation}\label{4.65}
\sum_{i=1}^n 
d_i(\bar{\alpha}^i)^2=1-\frac{1}{q^2}\frac{\kappa^2A^{(1)}}
{\eps+\kappa^2A^{(1)}}\rightarrow 1\quad \mbox{for} \quad A^{(1
)}\rightarrow 0
\end{equation}
in accordance with the Eqs. \rf{3.40} to \rf{3.42}. For the 
volume of the universe, we obtain in this limit
\begin{equation}\label{4.66}
V\sim t,\quad t\rightarrow 0.
\end{equation}
Thus, for positive energy, $E>0$, and small synchronous times the universe 
behaves like the Kasner universe. 

Now, we consider the exceptional case $E=0$. It follows from \rf{4.12} that
\begin{equation}\label{4.67}
\tau\approx \frac{2}{q^2B}\frac{1}{t},\quad t\rightarrow 0
\end{equation}
and for the volume, we obtain from \rf{4.15}
\begin{equation}\label{4.68}
V\sim t^2,\quad t\rightarrow 0.
\end{equation}
With the help of Eq. \rf{3.16}, we conclude that the 
approximation of the scale factors is given by
\begin{equation}\label{4.69} 
a_i\approx 
\bar{A}_it^{2/(D-1)}\exp\left(\frac{2\alpha_i}{q^2B}\frac{1}{t}\right),
\quad t\rightarrow 0
\end{equation}
in accordance with expression \rf{3.44} for $\alpha^{(2)}=1$.

Thus, the scale factors behave either anisotropically and 
exponentially like
\begin{equation}\label{4.70}
a_i\sim \exp\left(\frac{2\alpha_i}{q^2B}\frac{1}{t}\right),\quad 
t\rightarrow 0
\end{equation}
if $\alpha_i\ne 0\quad (\eps>0,\, A^{(1)}<0)$ or they have 
power law behaviour like
\begin{equation}\label{4.71}
a_i\sim t^{2/(D-1)},\quad t\rightarrow 0
\end{equation}
if $\alpha_i=0\quad (\eps=0,\, A^{(1)}=0)$ (see Eq. \rf{3.45} for 
$\alpha^{(2)}=1$). 
In 
the 
latter 
case, the free minimally coupled scalar field is absent 
$(A^{(1)}=0)$.

Similar investigations can be done for the equation \rf{4.51} 
shifted in time such that $V\approx t^2,\quad t\rightarrow 
0$.

If $E<0$, the universe has in the Lorentzian region a 
classical turning point at the minimal volume $V_{min}$ and 
reaches never $V=0$ (see Fig. 1 and Fig. 2).

Now, let us consider the infinite volume limit $V\rightarrow 
0$ which, as we shall see, coincides with the limit 
$t\rightarrow +\infty$. As follows from Fig. 1 and Fig. 2, 
this is possible if $B,\, C>0$ or $B<0,\, C>0$. With 
Eqs. \rf{4.12} to \rf{4.14} and \rf{4.35}, \rf{4.37} to \rf{4.39} one 
can demonstrate that $\tau$ becomes asymptotically constant for
$t\rightarrow +\infty$ and the constant can be put equal to 
zero (with a proper choice of the integration constant 
$\tau_0$). From \rf{3.16} follows that isotropization takes 
place in this limit, namely
\begin{equation}\label{4.72}
a_i\sim V^{1/D-1},\quad t\rightarrow +\infty
\end{equation}
and from Eqs. \rf{4.15} to \rf{4.17} and \rf{4.40}, \rf{4.42} to 
\rf{4.44} we get
\begin{equation}\label{4.73}
V\approx \exp\left(\sqrt{2C}qt\right),\quad t\rightarrow 
+\infty
\end{equation}
in accordance with \rf{3.33}.

Thus, if $C>0$ we obtain in the limit $t\rightarrow +\infty$ 
an (isotropic) de Sitter universe. If $C<0$, the universe 
has a classical turning point at maximal volume $V_{max}$ 
and the volume can not go to infinity.

Let us come back once more to the case $C>0$ describing a 
universe arising from ``nothing". The volume is given by 
\rf{4.20}, \rf{4.44}, and \rf{4.42} and the harmonic time coordinate 
and the synchronous one are related via \rf{4.19}, \rf{4.39}, and 
\rf{4.37}, respectively. We shall restrict ourselves to the 
case $E<0$ for simplicity. In this case, we get the asymptotic
expression
\begin{equation}\label{4.74}
\tau\approx 
\frac{1}{\sqrt{|E|/2}q}\arctan\frac{\sqrt{2|E|C}}{\sqrt{|\Delta|}-B}\equiv 
A,
\end{equation}
if $t\gg \left(\sqrt{\frac{C}{2}}q\right)^{-1}$ (it is sufficient to 
satisfy $\sqrt{\frac{C}{2}}qt\ge 2$). Then, as follows from equation 
\rf{3.16}, the scale factors are given by
\begin{equation}\label{4.75}
a_i\approx A_i\exp(\alpha^iA)V^{1/(D-1)}.
\end{equation}
In \cite{11} was shown that for
$4{\lorder}\sqrt{\frac{C}{2}}qt\ll D-1$ the parameters of the 
model can be chosen such that, due to the exponential 
prefactor in \rf{4.75}, some of the factor spaces (with $\alpha^i>0$)
 undergo inflation 
after birth from ``nothing" while other factor spaces (with 
$\alpha^i<0$) remain  compactified near the Planck length 
$L_{\Pl}\approx 10^{-33}cm$. The (graceful exit) mechanism 
responsible for the transition 
from the inflationary stage to the Kasner-like stage, in 
which the scale factors of the external spaces $M_i$ exhibit 
power-law behaviour while the internal spaces remain frozen 
in near the Planck scale,   
deserves still more detailed investigations, similar those of \cite{11}.
(There the complementary case 
of multidimensional cosmological models with cosmological 
constant was considered.)
%
\nl
\vspace*{0.8cm}
\nl\noindent
{\bf 5. CLASSICAL WORMHOLES}
\vspace*{0.4cm}
\nl\noindent
In this chapter we consider in more detail a special
type of instantons, called wormholes.
These usually are Riemannian metrics, consisting
of two large regions joined by a narrow throat (handle).
Obviously, they appear if the classical Lorentzian solutions
of the model have turning points at some maximum,
Namely, according to Fig. 1 and 2, for models with $C<0$
(the parameter $B$ may be positive as well as negative).
Let us show this explicitly. We consider instantons
which can be obtained by analytic continuation $t\to -it$ of the
Lorentzian solutions \rf{4.51},  \rf{4.52}, \rf{4.57} and \rf{4.60}
respectively.
\begin{eqnarray}\label{5.1}
V&=&\frac{B}{|C|}\cosh^2\left(\sqrt{|C|/2}qt\right) ,\quad 
E=0,\quad |t|< \infty ,
\\\label{5.2}
V&=&\frac{1}{2|C|}
\left[\sqrt{|\Delta|}+B
+\left(\sqrt{|\Delta|}-B\right)\tanh^2\left(\sqrt{|C|/2}qt\right)\right] 
\cosh^2\left(\sqrt{|C|/2}qt\right),
\\\nonumber
&&E>0,\quad |t|< \infty , 
\\\label{5.3}
V&=&\frac{1}{2|C|}\left[B+\sqrt{|\Delta|}
-\left(B-\sqrt{|\Delta|}\right)\tanh^2\left(\sqrt{|C|/2}qt\right)\right]
\cosh^2\left(\sqrt{|C|/2}qt\right),
\\\nonumber
&&-B^2/4|C|<E<0,\quad |t|< \infty ,
\\\label{5.4}
V&=&\frac{1}{2|C|}
\left[\sqrt{|\Delta|}-|B|
+\left(\sqrt{|\Delta|}+|B|\right)
\tanh^2\left(\sqrt{|C|/2}qt\right)\right]
\cosh^2\left(\sqrt{|C|/2}qt\right),
\\\nonumber
&&E>0,\quad |t|< \infty . 
\end{eqnarray}
As mentioned before, these equations correspond (with evident substitutions)
to the Lorentzian equations 
\rf{4.40},  \rf{4.44}, \rf{4.42} and \rf{4.20}, respectively.
The harmonic and synchronous times are related respectively by
\begin{eqnarray}\label{5.5}
\tau=\frac{\sqrt{2|C|}}{Bq}\tanh\left(\sqrt{|C|/2}qt\right), 
&& E=0 ,
\\\label{5.6}
\tan\left(\sqrt{E/2}q\tau\right)=\frac{\sqrt{4E|C|}}{\sqrt{|\Delta|}+B}
\tanh\left(\sqrt{|C|/2}qt\right),
&& E>0,
\\\label{5.7}
\tanh\left(\sqrt{|E|/2}q\tau\right)=\frac{\sqrt{|\Delta|}-B}{\sqrt{4EC}}
\tanh\left(\sqrt{|C|/2}qt\right),
&& -\frac{B^2}{4|C|}<E<0  , 
\\\label{5.8}
\tan\left(\sqrt{E/2}q\tau\right)=\frac{\sqrt{4E|C|}}{\sqrt{|\Delta|}-|B|}
\tanh\left(\sqrt{|C|/2}qt\right),
&& E>0 . 
\end{eqnarray}
(See  \rf{4.35}, \rf{4.39}, \rf{4.37} and \rf{4.19} respectively.  
\rf{4.37} looks like \rf{5.7}, 
if we choose the constant of integration $\tau_0$ 
such that $f|_{\tau=0}=V_0^{(2)}$, 
whence $f=\sqrt{|\Delta|} \exp\left(\sqrt{2E}q\tau\right)$, 
and use the relation $f=|B|-{\sqrt{4EC}}
\tanh\left[\sqrt{C/2}qt
+\artanh\frac{|B|-\sqrt{|\Delta|}}{\sqrt{4EC}}\right]$, 
where a turning point appears for $t=0$.)

It can easily be seen from  \rf{5.5} to \rf{5.8}
that the harmonic time $\tau$ is finite for the full range
$-\infty<t<\infty$ and goes to constants when $|t|\to+\infty$.

For the spatial volume of the universe we have the asymptotic behaviour
\be{5.9}
V \sim  \exp\left({\sqrt{2|C|}\,q\,|t|}\right) , \quad  |t|\to\infty ,    
\ee
for all cases \rf{5.1} to \rf{5.4}.

In the Euclidean region   \rf{3.16} holds unchanged,
since the Wick rotation $\tau\to-i\tau$ has to be accompanied
by the transformation $\alpha^j\to i \alpha^j$ ($p^j\to i p^j$). 
This means that the parameter $\eps$ in the constraint equation \rf{3.12}
remains unchanged (see \rf{3.11}).

Thus, the Euclidean metric reads
\be{5.10}
ds^2=dt^2 + a_1^2(t)g^{(1)}+\ldots+a_n^2(t)g^{(n)},
\ee
where each scale factor $a_i$ has its own turning point at "time" $t_i$,
when $\frac{d}{dt}a_i=0$.
The metric has its most simple and symmetric form in the case 
$\eps=0$ ($\alpha^i=0$, $i=1,\ldots,n$), whence 
\be{5.11}
ds^2=dt^2 + V^{\frac{2}{D-1}}\left(g^{(1)}+\ldots+g^{(n)}\right),
\ee
where $V$ is given by equations \rf{5.1} to \rf{5.4}, and the throat
is located at $t=0$.
In the limit $|t|\to\infty$, the metric \rf{5.10} converges to
\be{5.12}
ds^2=dt^2 + \exp\left(\frac{2\sqrt{2|C|}}{D-1}\,q\,|t|\right)
\left(g^{(1)}+\ldots+g^{(n)}\right) ,
\ee
which describes an asymptotically anti-de Sitter Euclidean universe.
Thus, the metric \rf{5.10} describes asymptotically
anti-de Sitter wormholes.

The structure of a universe for models with classical
Euclidean wormholes is shown schematically in Fig. 5 und Fig. 6 
for the symmetric case $\alpha^i=0$ ($i=1,\ldots,n$) with a metric
\rf{5.11}. (Note that, 
\rf{5.11} has a common scale factor for all factor spaces,
while for the universe of Eq. (87) in \cite{KZ}, 
sketched by similar looking Fig. 3 resp. Fig. 4 there,   
internal spaces are static, but external space is not.)
There are two qualitatively different pictures. 
The first case (see Fig. 5) takes place for $E\ge 0$ ($B>0$, $C<0$)  
and for $E>0$ ($B,C<0$) and describes asymptotically
an anti-de Sitter wormhole and a baby universe which can branch off 
from this wormhole.
The second case (see Fig. 6) takes place for $-B^2/4|C|<E< 0$ ($B>0$, $C<0$)  
and describes, besides
wormhole and baby universe, an additional parent instanton which is
responsible for the birth of the universe from nothing.
\nl
\vspace*{0.8cm}
\nl\noindent
{\bf 6. RECONSTRUCTION OF THE POTENTIALS}
\vspace*{0.4cm}
\nl\noindent
The effective perfect fluid Lagrangian \rf{2.11} has its origin 
in the scalar field Lagrangian \rf{2.6}. In this Chapter, we 
calculate a class of potentials 
$U^{(a)}(\varphi^{(a)})$ which ensure the equivalence of these Lagrangians.  
The 
procedure of potential reconstruction was proposed in \cite{Zhuk} 
and is applied in the following. For the 
integrable 3-component model holds
\begin{equation}\label{6.1}
\varphi^{(a)}=\pm \frac{\sqrt{\alpha^{(a)}/2}}{q}\int
\frac{\sqrt{\rho^{(a)}(V)}dV}{\left[\eps+\kappa^2V^2\left(\rho^{(1)}
+\rho^{(2)}+\rho^{(3)}\right)\right]^{1/2}}+\varphi_0^{(a)},\quad
a=1, 2, 3,
\end{equation}
where the energy density $\rho^{(a)}$ is given by \rf{2.12} and 
$\alpha^{(1)}=2$, $\alpha^{(2)}=1$, $\alpha^{(3)}=0$. 
$\varphi_0^{(a)}$ is a constant of integration. We should 
stress that equation \rf{6.1} was obtained for Lorentzian 
regions. As a result, we get the scalar fields 
$\varphi^{(a)}$ as a function of the spatial volume V. 
Inverting this expression, we find the spatial volume as a 
function of the scalar field $\varphi^{(a)}$, 
$V=V(\varphi^{(a)})$, and consequently, a dependence of the 
energy density $\rho^{(a)}$ on the scalar field 
$\varphi^{(a)}$, $\rho^{(a)}=\rho^{(a)}(\varphi^{(a)})$. Then, 
using Eqs. \rf{2.8} to \rf{2.10}, we find the potential 
$U^{(a)}(\varphi^{(a)})$ in the form
\begin{equation}\label{6.2}
U^{(a)}(\varphi^{(a)})=\frac{1}{2}
\left(2-\alpha^{(a)}\right)\rho^{(a)}(\varphi^{(a)}), \quad
a=1, 2, 3,
\end{equation}
where
\begin{equation}\label{6.3}
\rho^{(a)}=A^{(a)}\left[V(\varphi^{(a)})\right]^{-\alpha^{(a)}}.
\end{equation}
The third component of the scalar field has 
$\alpha^{(3)}=0$. Then, from \rf{6.2} and \rf{6.3} it follows that 
$\varphi^{(3)}$, $U^{(3)}$, and $\rho^{(3)}$ are constant. 
This scalar field component with the equation of state 
$P^{(3)}=-\rho^{(3)}$ is equivalent to the 
cosmological constant $\Lambda \equiv 
\kappa^2U^{(3)}=\kappa^2A^{(3)}=C$. For $\alpha^{(1)}=2$, we 
have $U^{(1)}\equiv 0$ (free scalar field). In this case, 
the scalar field $\varphi^{(1)}$ is equivalent to a
ultra-stiff perfect fluid ($P^{(1)}=\rho^{(1)}$). Equation 
\rf{6.1} reads in this case
\begin{equation}\label{6.4}
\varphi^{(1)}-\varphi_0^{(1)}=\pm\frac{\sqrt{A^{(1)}}}{q}
\int\frac{dV}{V\sqrt{E+BV+CV^2}},
\end{equation}
where $E$ and $B$ are defined by \rf{4.3} and \rf{4.4} respectively.

A consequence of \rf{4.5} is
\begin{equation}\label{6.5}
\varphi^{(1)}-\varphi_0^{(1)}=\pm\sqrt{2A^{(1)}}\tau.
\end{equation}
This result is expected for a free minimal coupled scalar 
field in the harmonic time gauge where 
$\ddot{\varphi}^{(1)}=0$. After integration in \rf{6.4},
\begin{eqnarray}\label{6.6}
\varphi^{(1)}-\varphi_0^{(1)}=
\mp \,i\, \frac{2\sqrt{|A^{(1)}|}}{q}
\frac{\sqrt{BV+CV^2}}{BV},
&& E=0,
\\\label{6.7}
\varphi^{(1)}-\varphi_0^{(1)}=
\pm \frac{\sqrt{A^{(1)}}}{\sqrt{E}q}\ln\frac{2E+BV
-2\sqrt{ER}}{2V},
&& E>0,
\\\label{6.8}
\varphi^{(1)}-\varphi_0^{(1)}=\pm \,i\, 
\frac{\sqrt{|A^{(1)}|}}{\sqrt{|E|}q}\arcsin\frac{BV-2|E|}
{V\sqrt{|\Delta|}},
&& E<0,\quad B^2-4EC>0,
\end{eqnarray}
with $R:=E+BV+CV^2$ and $|\Delta|=B^2-4EC$.
For $E\leq 0$ (i.e. $A^{(1)}<0$) this scalar field is imaginary.

Let us now consider the second component with $\alpha^{(2)}=1$.
The scalar field $\varphi^{(2)}$ is equivalent to dust 
($P^{(2)}=0$). Equation \rf{6.1} reads now
\begin{equation}\label{6.9}
\varphi^{(2)}-\varphi_0^{(2)}=\pm\frac{\sqrt{A^{(2)}}}{\sqrt{2}q}
\int\frac{dV}{\sqrt{V}\sqrt{E+BV+CV^2}}.
\end{equation}
$\varphi^{(2)}$ is imaginary for $A^{(2)}<0$, i.e. $B<0$. The 
integral in \rf{6.9} is an elliptic one and, in general, it is 
not possible to express it by elementary functions. But in 
the particular case $E=0$, which expresses the  
asymptotic behaviour of  the scalar field \rf{6.9}, we get
\begin{eqnarray}\label{6.10}
\varphi^{(2)}-\varphi_0^{(2)}=\mp \frac{\sqrt{2}}{\kappa q}
\arcoth\left(1+\frac{C}{B}V\right)^{1/2},
&& B, C > 0,
\\\label{6.11}
\varphi^{(2)}-\varphi_0^{(2)}=\mp \frac{\sqrt{2}}{\kappa q}
\artanh\left(1-\frac{|C|}{B}V\right)^{1/2},
&& B>0,\ C<0,
\\\label{6.12}
\varphi^{(2)}-\varphi_0^{(2)}=\mp i \frac{\sqrt{2}}{\kappa q}
\arctan\left(\frac{C}{|B|}V-1\right)^{1/2},
&& B<0,\ C>0,
\end{eqnarray}
the volume of the universe
\begin{eqnarray}\label{6.13}
V=\frac{B}{C}\sinh^{-2}\left[\frac{\kappa 
q}{\sqrt{2}}\left(\varphi^{(2)}-\varphi_0^{(2)}\right)\right], 
&& B, C>0,
\\\label{6.14}
V=\frac{B}{|C|}\cosh^{-2}\left[\frac{\kappa 
q}{\sqrt{2}}\left(\varphi^{(2)}-\varphi_0^{(2)}\right)\right], 
&& B>0,\ C<0,
\\\label{6.15}
V=\frac{|B|}{C}\cos^{-2}\left[\frac{\kappa 
q}{\sqrt{2}}i\left(\varphi^{(2)}-\varphi_0^{(2)}\right)\right], 
&& B<0,\ C>0,
\end{eqnarray}
and the potential of the scalar field
\begin{eqnarray}\label{6.16}
U^{(2)}(\varphi^{(2)})=\frac{C}{2\kappa^2}\sinh^2\left[\frac{\kappa 
q}{\sqrt{2}}\left(\varphi^{(2)}-\varphi^{(2)}_0\right)\right],
&& B, C>0,
\\\label{6.17}
U^{(2)}(\varphi^{(2)})=\frac{|C|}{2\kappa^2}\cosh^2\left[\frac{\kappa 
q}{\sqrt{2}}\left(\varphi^{(2)}-\varphi^{(2)}_0\right)\right],
&& B>0,\ C<0,
\\\label{6.18}
U^{(2)}(\varphi^{(2)})=\frac{C}{2\kappa^2}\cos^2\left[\frac{\kappa 
q}{\sqrt{2}}i\left(\varphi^{(2)}-\varphi^{(2)}_0\right)\right],
&& B<0,\ C>0.
\end{eqnarray}
It follows from these equations that for $B, C>0$ and $B, 
C<0$ the volume goes to infinity like
\begin{equation}\label{6.19}
V\sim \frac{1}{\left|\varphi^{(2)}\right|^2}\to +\infty,\quad  
\left|\varphi^{(2)}\right|\to 0.
\end{equation}
The general expression \rf{6.9} should have the same asymptotic 
behaviour in all the cases where the limit $V\to +\infty$ is 
permitted, because we can drop in this limit the term E in 
the denominator of \rf{6.9}.

If $E>0$, from \rf{6.9} results
\begin{equation}\label{6.20}
\varphi^{(2)}-\varphi^{(2)}_0\approx 
\frac{\sqrt{2A^{(2)}/E}}{q}\sqrt{V}\to 0,\quad V\to 0.
\end{equation}
Let us now consider two particular cases of \rf{6.9} for 
$E\ne 0$. The first case is that one when the classical trajectory 
has two turning points $V_0^{(1,2)}$, i.~e. when either 
$B>0,\, C<0$ or $B<0,\, C>0$ ( for both the cases 
$B^2>4EC$). Then (see equation (3.131) in \cite{12}), 
\begin{equation}\label{6.21}
\varphi^{(2)}=\pm \frac{2}{\kappa 
q}\frac{\sqrt{B}}{\sqrt{|B|+\sqrt{|\Delta|}}}F(\psi|m)
\end{equation}
where $F(\psi|m)$ is the elliptic integral of the first 
kind \cite{13} and
\begin{eqnarray}\label{6.22}
\psi=\arcsin\sqrt{(V_0^{(2)}-V)/(V_0^{(2)}-V_0^{(1)})},
&&
V_0^{(2)}>V\ge V_0^{(1)},
\quad B>0, C<0,
\\\label{6.23}
\psi=\arcsin\sqrt{V/V_0^{(1)}},
&& V_0^{(2)}>V_0^{(1)}\ge V,
\quad B<0,\ C>0,
\\\label{6.24}
\psi=\arcsin\sqrt{(V-V_0^{(2)})/(V-V_0^{(1)})},
&& V>V_0^{(2)}>V_0^{(1)},
\quad B<0,\ C>0,
\end{eqnarray}
\begin{eqnarray}\label{6.25}
m=\sqrt{1-V_0^{(1)}/V_0^{(2)}},
&& V_0^{(2)}>V\ge V_0^{(1)},
\quad B>0,\ C<0,
\\\label{6.26}
m=\sqrt{V_0^{(1)}/V_0^{(2)}},
&& V_0^{(2)}>V_0^{(1)}\ge V, 
\quad B<0,\ C>0,
\\\label{6.27}
m=\sqrt{V_0^{(1)}/V_0^{(2)}},
&& V>V_0^{(2)}>V_0^{(1)},
\quad B<0,\ C>0.
\end{eqnarray}
The turning points are
\begin{equation}\label{6.28}
V_0^{(1, 2)}=\frac{|B|\mp \sqrt{|\Delta|}}{2|C|}.
\end{equation} 
The minus sign is related to $V_0^{(1)}$, the plus sign to 
$V_0^{(2)}$, and $|\Delta| =B^2-4EC$. 
The scalar field $\varphi^{(2)}$ is imaginary for $B<0$. 

With the Jacobian elliptic functions \cite{13}, inverting \rf{6.21}, the 
volume of the universe is given by
\begin{eqnarray}\label{6.29}
V=V_0^{(2)}- \sn^2 (V_0^{(2)}-V_0^{(1)}),
&& 
V_0^{(2)}>V\ge V_0{(1)},
\quad B>0,\ C<0,
\\\label{6.30}
V=\sn^2 V_0^{(1)},
&& V_0^{(2)}>V_0^{(1)}\ge V,
\quad B<0,\ C>0,
\\\label{6.31}
V=\frac{V_0^{(2)}-\sn^2 V_0^{(1)}}{1-\sn^2},
&& 
V>V_0^{(2)}>V_0^{(1)},
\quad B<0,\ C>0,
\end{eqnarray}
where $\sn\equiv \sn\left(I_1\varphi^{(2)}|m\right)=\sin\psi$ and
$I_1^{-1}=\pm \frac{2\sqrt{B}}{\kappa q\sqrt{|B|+|\Delta|}}$. The 
corresponding potential terms are then given as 
$U^{(2)}=A^{(2)}/2V$ (see \rf{6.2} and \rf{6.3} for 
$\alpha^{(2)}=1$). According to the properties of the Jacobian elliptic 
functions \cite{13} asymptotic estimates for \rf{6.30} and \rf{6.31} are 
$V\approx (q^2E/2A^{(2)})(\varphi^{(2)})^2$ for $|\varphi^{(2)}|\to 0$ (in 
accordance with \rf{6.20}) and $V\sim 1/|\varphi^{(2)}|^2$ for 
$|\varphi^{(2)}|\to 0$ (in accordance with \rf{6.19}).

Another particular case for $C>0$ is that with $E>B^2/4C$. 
Here (see (3.138 (7)) in \cite{12}), we obtain 
\begin{equation}\label{6.32}
\varphi^{(2)}=\pm \frac{\sqrt{B/2C}}{\kappa 
q}\frac{1}{(E/C)^{1/4}}F(\psi|m),
\end{equation}
where
\begin{equation}\label{6.33}
\psi =2\arctan\sqrt{V/\sqrt{E/C}}
\end{equation}
and
\begin{equation}\label{6.34}
m=\sqrt{(2\sqrt{EC}-B)/4\sqrt{EC}}.
\end{equation}
Inverting equation \rf{6.32}, we obtain
\begin{equation}\label{6.35}
V=\frac{2-\sn^2}{\sn^2}\sqrt{E/C}\pm
\sqrt{\left(\frac{2-\sn^2}{\sn^2}\sqrt{E/C}\right)^2-\frac{E}{C}}
\end{equation}
with $\sn\equiv \sn(I_2\varphi^{(2)}|m)=\sin \psi$ and 
$I_2^{-1}=\pm \frac{\sqrt{B/2C}}{\kappa q(E/C)^{1/4}}$. For the 
branch with the plus sign, $V\sim \frac{1}{|\varphi^{(2)}|^2}\to \infty$ 
for $|\varphi^{(2)}|\to 0$ (in accordance with \rf{6.19}) 
and for the branch with the minus sign, 
$V\approx (q^2E/2A^{(2)})(\varphi^{(2)})^2\to 0$
for $|\varphi^{(2)}|\to 0$ (in accordance with \rf{6.20}). 
To find the scalar field potential, we have to substitute 
\rf{6.35} into $U^{(2)}(\varphi^{(2)})=A^{(2)}/2V$.
\nl
\vspace*{0.8cm}
\nl\noindent
{\bf 7. SOLUTIONS TO THE QUANTIZED MODEL}
\vspace*{0.4cm}
\nl\noindent
At the quantum level, the constraint equation \rf{3.8} is 
replaced by the Wheeler-DeWitt (WDW) equation. The 
WDW equation is covariant with respect to gauge as well as minisuperspace 
coordinate transformations \cite{Ra}. In the 
harmonic time gauge \cite{IMZ,Ra} it reads
\begin{equation}\label{7.1}
\left(\frac{1}{2}\frac{\partial^2}{\partial {z^0}^2}-\frac{1}{2}
\sum_{i=1}^{n-1}\frac{\partial^2}{\partial {z^i}^2}+
\kappa^2\sum_{a=1}^{m}A^{(a)}\exp(k^{(a)}qz^0)\right)\Psi=0.
\end{equation}
Formally, this WDW equation has the same structure as 
that of \cite{KZ}.  
However, on the semiclassical level the dynamics of the 
universe is quite different for the models in both the 
papers. Semiclassical equations were considered in \cite{15}. 

We look for solutions of \rf{7.1} by separation of the 
variables and try the ansatz
\begin{equation}\label{7.2}
\Psi(z)=\Phi(z^0)\exp\left(i{\p}\cdot{\z}\right),
\end{equation}
where ${\p}:=(p^1,\ldots,p^{n-1})$ is a constant vector, 
${\z}:=(z^1,\ldots,z^{n-1})$, $p_i=p^i$ and
${\p}\cdot{\z}:=\sum_{i=1}^{n-1}p_iz^i$. Substitution of 
\rf{7.2} into \rf{7.1} gives
\begin{equation}\label{7.3}
\left[\frac{1}{2}\frac{d^2}{d{z^0}^2}+\frac{1}{2}\sum_{i=1}^{n-1}(p_i)^2
+\kappa^2\sum_{a=1}^{m}A^{(a)}\exp\left(k^{(a)} q z^0\right)\right]\Phi=0.
\end{equation}
For the integrable 3-component model this equation reduces 
to 
\begin{equation}\label{7.4}
\left[-\frac{1}{2}\frac{d^2}{d{z^0}^2}+U(z^0)\right]\Phi=E\Phi
\end{equation}
in the notation of \rf{4.3} and \rf{4.4}. Following \cite{KZ}, we 
rewrite this equation like
\begin{equation}\label{7.5}
x^2\frac{d^2\Phi}{dx^2}+x\frac{d\Phi}{dx}+
\left[\tilde{E}+\tilde{B}x+\tilde{C}x^2\right]\Phi=0,
\end{equation}
where $\tilde{E}=2E/q^2$, $\tilde{B}=2B/q^2$, 
$\tilde{C}=2C/q^2$, and $x=\exp(qz^0)$ ($x$ is identical to the spatial 
volume 
of the universe V). \rf{7.5} is equivalent to the Whittaker 
equation
\begin{equation}\label{7.6}
\frac{d^2y}{d\xi^2}+\left[-\frac{1}{4}+\frac{\tilde{B}/T}{\xi}+
\frac{\tilde{E}+1/4}{\xi^2}\right]y=0,
\end{equation}
where $T:=\pm 2i\sqrt{\tilde{C}}$, $\xi:=Tx$, and 
$\Phi=:x^{-1/2}y(\xi)$ and also equivalent to the Kummer 
equation
\begin{equation}\label{7.7}
\xi\frac{d^2w}{d\xi^2}+(1+2\mu-\xi)\frac{dw}{d\xi}-
\left[\frac{1}{2}+\mu-\frac{\tilde{B}}{T}\right]w=0,
\end{equation}
where $\mu^2:=-\tilde{E}$ and 
$\Phi=:x^{-1/2}\exp\left(-\frac{1}{2}\xi\right)\xi^{\frac{1}{2}+\mu}w(\xi)$.
In the first case, the solutions are the Wittaker functions 
\cite{13} $y_1:=M_{k,\mu}(\xi)$ and $y_2:=W_{k,\mu}(\xi)$ with 
$k:=\tilde{B}/T$ and $\mu^2:=-\tilde{E}$. In the second case, 
the solutions are the Kummer functions \cite{13} $w_1:=M(a,b,\xi)$ 
and $w_2:=U(a,b,\xi)$ with $a:=\frac{1}{2}+\mu-\tilde{B}/T$ 
and $b:=1+2\mu$.

The general solution of equation \rf{7.1} for the 3-component 
model is
\begin{equation}\label{7.8}
\Psi(z)=\sum_{i=1,2}\int d^{n-1}p\, C_i({\p})\exp\left(i{\p}\cdot{\z}\right)\,
\Phi^{(i)}_E(\exp(qz^0)),
\end{equation}
where  
$\Phi^{(1,2)}_E=\frac{1}{\sqrt{x}} y_{1,2}(\xi)$, or
$\Phi^{(1,2)}_E=\frac{1}{\sqrt{x}} \exp\left(-\frac{1}{2}\xi\right)
\xi^{\frac{1}{2}+\mu}w_{1,2}(\xi)$.
It is convenient to set $T=+2i\sqrt{\tilde{C}}$ for $C>0$ 
and $T=-2i\sqrt{\tilde{C}}$ for $C<0$, and 
$\mu:=+\sqrt{-\tilde{E}}$.

In \cite{15}, it was argued that the parameter $E$ can be 
interpreted as energy. So, the state with $E=0$, vanishing 
momenta $p_i\, (i=1,\ldots,n-1)$, and $A^{(1)}=0$ (absence of 
free scalar field excitations) is the 
ground state of the system. Thus, its wave function reads
\begin{equation}\label{7.9}
\Psi_0=\Phi_0^{(i)}\left(\exp(qz^0)\right),\quad i=1,2.
\end{equation}
The limit of large spatial geometries in \rf{7.5} $z^0\to 
+\infty$ (remember $x\equiv V=\exp(qz^0)$) is equivalent to 
$\tilde{B}\to 0$. In this limit, the Wittaker functions 
reduce to Bessel functions \cite{13}, namely
\begin{equation}\label{7.10}
M_{k,\mu}(\xi)\mathop{\longrightarrow}\limits_{k\to 0} 
\sqrt{V}J_{\mu}\left(\sqrt{\tilde{C}}V\right),
\end{equation}
\begin{equation}\label{7.11}
W_{k,\mu}(\xi)\mathop{\longrightarrow}\limits_{k\to 0} 
\sqrt{V}H_{\mu}^{(2)}\left(\sqrt{\tilde{C}}V\right)
\end{equation}
for $C>0$ and
\begin{equation}\label{7.12}
M_{k,\mu}(\xi)\mathop{\longrightarrow}\limits_{k\to 0} 
\sqrt{V}I_{\mu}\left(\sqrt{|\tilde{C}|}V\right),
\end{equation}
\begin{equation}\label{7.13}
W_{k,\mu}(\xi)\mathop{\longrightarrow}\limits_{k\to 0} 
\sqrt{V}K_{\mu}\left(\sqrt{|\tilde{C}|}V\right)
\end{equation}
for $C<0$.

Following the ideas of \cite{15,16}, one can demonstrate that 
for $C>0$ the wave function
\begin{equation}\label{7.14}
\Psi_0^{HH}=\Phi_0^{(1)}\mathop{\longrightarrow}\limits_{k\to 
0}J_0\left(\frac{\sqrt{2C}}{q}V\right)\mathop{\sim}\limits_{V\to 
\infty}\cos\left(\frac{\sqrt{2C}}{q}V\right)
\end{equation}
corresponds to the Hartle-Hawking boundary condition \cite{17} 
and the wave function
\begin{equation}\label{7.15}
\Psi_0^{V}=\Phi_0^{(2)}\mathop{\longrightarrow}\limits_{k\to 
0}H_0^{(2)}\left(\frac{\sqrt{2C}}{q}V\right)\mathop{\sim}\limits_{V\to 
\infty}\exp\left(-i\frac{\sqrt{2C}}{q}V\right)
\end{equation}
corresponds to the Vilenkin boundary condition \cite{18}.

In the case $C<0$, we get
\begin{equation}\label{7.16}
\Psi_0^{HH}=\Phi_0^{(1)}\mathop{\longrightarrow}\limits_{k\to 
0}I_0\left(\frac{\sqrt{2|C|}}{q}V\right)\mathop{\sim}\limits_{V\to 
\infty}\exp\left(\frac{\sqrt{2|C|}}{q}V\right)
\end{equation}
The potential has for $B>0,\,C<0$ a well and for $E<0$ the 
energy spectrum is discrete (see Fig. 2). In this case, the finite 
solutions of the wave equation \rf{7.4} \cite{19} are
\begin{equation}\label{7.17}
\Phi_n=\exp\left(-\frac{1}{2}\xi\right)\xi^{\mu}M(-n,b,\xi).
\end{equation}
The energy levels are given by
\begin{equation}\label{7.18}
-E_n=\left[\frac{B}{2\sqrt{|C|}}-\frac{q}{\sqrt{2}}
\left(n+\frac{1}{2}\right)\right]^2.
\end{equation}
$n$ is a non negative integer and restricted to
\begin{equation}\label{7.19}
n<\frac{B}{q\sqrt{2|C|}}-\frac{1}{2}.
\end{equation}
Thus, the discrete spectrum has a finite number of 
eigenvalues. 
If $\frac{B}{q\sqrt{2|C|}}<\frac{1}{2}$, there is no discrete 
spectrum.
It was demonstrated in \cite{KZ} that the wave functions \rf{7.17} 
satisfy the quantum wormhole boundary conditions \cite{20}.
%
%
\nl
\vspace*{0.8cm}
\nl\noindent
{\bf 8. CONCLUSIONS}
\vspace*{0.4cm}
\nl\noindent
We considered the generalization of a homogeneous cosmological model of
Bianchi type I to an anisotropic multidimensional one with $n\geq 2$
Ricci-flat spaces of arbitrary dimensions, in the presence of
$m$ homogeneous non-interacting minimally coupled scalar fields.
Under certain conditions these models are equivalent to multidimensional 
cosmological models in the presence of an $m$-component perfect fluid
with equations of state 
$P^{(a)}= \left( \alpha^{(a)} - 1\right)\rho^{(a)}$
with matter constants $\alpha^{(a)}$ for $a=1,\ldots,m$.
Using this equivalence, for $m=3$, we find integrable models
when one of the scalar fields is equivalent to an 
ultra-stiff perfect fluid component, the second one corresponds
to dust, and the third one is equivalent to a vacuum component.
Dynamics of the universe was investigated in general, as well as 
in a particular $3$-component integrable case. For integrable models,
there are four qualitatively different types of evolution
of the universe, depending  on the potential $U(z_0)$
(see Fig. 1 and Fig. 2), but in all four cases the universe
has a Kasner-like behaviour near the cosmological singularity
(see \ref{4.62}).
In the cases where the universe can expand to infinity,
an isotropization takes place which results in an
asymptotically de Sitter universe (see  \rf{4.72} and  \rf{4.73}).

In quantum cosmology, instantons, solutions of the classical
Einstein equations in Euclidean space, play an important role,
giving significant contributions to the path integral. They are
connected with the changing geometry of the model.
We found here three interesting types of instantons.
The first one (see \rf{4.46} and \rf{4.48}) describes
tunnelling between a Kasner-like universe and an asymptotically
de Sitter universe. Sewing a number of these instantons
(see Fig. 4) may provide the Coleman mechanism for
the vanishing of the cosmological constant.
Another type of instanton (see  \rf{4.21} and \rf{4.47})
is responsible for the birth of the universe from "nothing".
It was shown that corresponding Lorentzian solutions \rf{4.20} and 
\rf{4.44}) can ensure inflation of the external space
(see \rf{4.75}) and compactification of the internal ones.
This problem needs a more detailed investigation in a separate paper.
The third type of instantons (see equations \rf{5.1} to \rf{5.4})
describes the Euclidean space which has an asymptotically anti-de Sitter 
wormhole geometry (see Fig. 5 and Fig. 6).

The scalar field potentials  $U^{(a)}(\varphi^{(a)})$ ($a=1,\ldots,m$)
can be reconstructed by the method described in \cite{Zhuk}.
We performed this procedure for integrable models, and exact
forms of potentials were presented in section 6.

The equivalence between a scalar field and a perfect fluid 
component helps also to investigate the quantum behaviour of
the universe. We obtained the Wheeler-de Witt equation 
from the effective perfect fluid Lagrangian.
Exact solutions are found, some of which describe cosmological transitions
with a signature change of the metric. e.g. universe nucleation
as quantum tunnelling from an Euclidean region.
Other solutions are given as quantum wormholes with discrete
spectrum (see \rf{7.18}).
\nl
{\bf Acknowledgements}
\nl
The work was supported by DFG grant 436~RUS~113/7/0. A.~Z. also thanks
the Projektgruppe
Kosmologie at the University Potsdam, as well as Prof. Kleinert and the 
Free University of Berlin for their hospitality during
preparation of this paper.
We also thank Ulrich Grimm for drawing the figures of this paper.
\nl
%

%
%
\newpage
{\Large
\noindent FIGURE CAPTIONS
}

\vspace{1cm}

\noindent Fig. 1. The potential $U_0(z^0)$ (solid line) and the energy
levels $E$ (dashed lines) in the cases $B,C>0$ and $B,C<0$.
Lorentzian regions exist for $E>U_0(z^0)$.

\vspace{1cm}

\noindent Fig. 2. The potential $U_0(z^0)$ (solid line) and the energy
levels $E$ (dashed lines)  in the cases $B>0$, $C<0$ and $B<0$, $C>0$.
In the former case we get a potential well, and in the latter case
we obtain a potential barrier.
Lorentzian regions exist for $E>U_0(z^0)$.
Here, $U_m=B^2/4C$ and $z_m^0=\frac{1}{q}\ln|B/2C|$.

\vspace{1cm}

\noindent Fig. 3. The qualitative shape of the instanton \rf{4.46}
(or \rf{4.48}).
The instanton describes tunnelling between a Kasner-like (baby) universe
and a de Sitter universe.

\vspace{1cm}

\noindent Fig. 4. Examples of instantons constructed by sewing
together several copies of the the instanton illustrated in Fig. 3.
Such instantons may describe tunnelling between 
(a) Kasner and de Sitter universe,
(b) two Kasner universes,
(c) two de Sitter universes.

\vspace{1cm}

\noindent Fig. 5. An asymptotically anti-de Sitter wormhole  is shown
schematically for energies $E\geq 0$ ($B>0$, $C<0$) and 
$E\geq 0$ ($B,C<0$) in the symmetrical case $\alpha^i=0$
($i=1,\ldots,n$). 
(Note that due to suppressed internal degrees of freedom
this looks like Fig. 3 in \cite{KZ},
although the solution and its context are quite different there.) 

\vspace{1cm}

\noindent Fig. 6. The qualitative structure of the universe is
is shown schematically for energies $-U_m<E< 0$ ($B>0$, $C<0$) 
in the symmetrical case $\alpha^i=0$
($i=1,\ldots,n$).
(Note that due to suppressed internal degrees of freedom
this looks like Fig. 4 in \cite{KZ},
although the solution and its context are quite different there.) 
 

\end{document}